\documentclass[aps,showpacs,nofootinbib]{revtex4}
\usepackage{epsf,latexsym}

\begin{document}

\title{On the statistical-mechanical meaning of the Bousso bound}
\author{Alessandro Pesci}
\email{pesci@bo.infn.it}
\affiliation
{INFN, Sezione di Bologna, Via Irnerio 46, I-40126 Bologna, Italy}

\begin{abstract}
The Bousso entropy bound, in its generalized form, 
is investigated
for the case of perfect fluids 
at local thermodynamic equilibrium
and evidence
is found that the bound is satisfied
if and only if a certain local 
thermodynamic property holds,
emerging when the attempt is made to apply the bound
to thin layers of matter.
This property
consists in the
existence 
of an ultimate lower limit $l^*$
to the thickness of the slices
for which a statistical-mechanical description
is viable, depending $l^*$ on the thermodynamical
variables which define the state of the system locally.
This limiting scale,
found to be in general much larger than the Planck scale
(so that no Planck scale physics must be necessarily
invoked to justify it), 
appears not related to gravity
and this suggests that the generalized entropy bound
is likely to be rooted on conventional flat-spacetime 
statistical mechanics, 
with the maximum admitted entropy
being however actually determined also 
by gravity.

Some examples of ideal fluids are considered
in order to identify the mechanisms
which can set a lower limit to the 
statistical-mechanical description
and these systems are found to respect
the lower limiting scale $l^*$.
The photon gas, in particular, appears to seemingly
saturate this limiting scale and the consequence is drawn 
that for systems consisting of a single
slice of a photon gas with thickness $l^*$,  
the generalized Bousso bound is saturated.
It is argued that
this seems to
open the way to a peculiar understanding
of black hole entropy:
if an entropy can meaningfully (i.e. with a second law)
be assigned to a black hole,
the value $A/4$ for it
(where $A$ is the area of the black hole) 
is required simply by
(conventional) statistical mechanics 
coupled to general relativity. 
\end{abstract}


\maketitle

$ $
According to 
the Bousso
covariant entropy bound \cite{Bousso1,Bousso2},
in a spacetime with Einstein equation
and obeying certain energy conditions, 
the entropy $S$ on a lightsheet $L$
(a lightlike hypersurface
spanned by non-expanding light rays emanating
orthogonally from 
an assigned 2-surface and
followed until they reach
a caustic or a singularity)
associated
to any given spacelike 2-surface $B$ with area $A$
satisfies

\begin{eqnarray}\label{BoussoBound}
S(L) \leq A/4
\end{eqnarray}
(in Planck units,
the units we will use
throughout this 
paper,
except when explicitly stated otherwise).
This bound, 
improving a previous formulation 
suggested in a cosmological context \cite{Fischler},
can be considered a covariant reformulation
of the bound \cite{tHooft1, Bekenstein94, tHooft2, Susskind}

\begin{eqnarray}\label{spherical}
S \leq A/4,
\end{eqnarray}
for the entropy $S$ inside a region
with boundary area $A$,
with the aim to overcome some inadequacies of 
the 
latter
and to obtain full general validity.
These bounds are intimately related to the so-called
holographic principle \cite{tHooft2, Susskind},
in broad terms
the conjecture that the physics 
of any spatial region is completely described 
by degrees of freedom living on its boundary
or also, in the spirit of the work of Bousso,
the physics of the lightsheets of any surface $B$
is completely described by degrees of freedom
living on $B$ \cite{Bousso2}.

A generalized form 
of bound (\ref{BoussoBound}) has been proposed 
in \cite{FlanaganMarolfWald};
it states that
if (some of) the light rays generating $L$ are terminated
on a 2-surface with area $A^\prime$
before reaching a caustic or singularity
so that $L$ is now a truncated lightsheet,
then

\begin{eqnarray}\label{GenBoussoBound}
S(L) \leq \frac{1}{4} (A-A^\prime)
\end{eqnarray}
and this clearly implies (\ref{BoussoBound})
as a particular case.
Proofs of the bound in this generalized form
can be found in \cite{FlanaganMarolfWald} itself 
and \cite{BoussoFlanaganMarolf,Strominger},
based on the individuation of 
some very general conditions
the material medium should obey, 
so that, if they hold,
the bound can be shown to be satisfied.  

In a recent paper \cite{Pesci} another proof has been given,
relying on Raychaudhuri equation evolved through
local Rindler frames (introduced in \cite{Jacobson}
as a way to unravel the thermodynamical meaning of
Einstein equation;
in \cite{Padmanabhan} also 
the relevance of these frames
for (horizon) thermodynamic
formulation of gravity in terms
of the surface term only of
Einstein-Hilbert action
is stressed).
This proof shows 
(actually for ideal fluids with $\mu \geq 0$,
where $\mu$, the chemical potential, 
includes the possible rest mass
of component particles)
that bound (\ref{GenBoussoBound})
is satisfied 
provided a certain inequality holds
(in flat spacetime, putting gravity aside),
which seems to set
a lower limit to the size 
of the systems for which a meaningful notion of 
statistical entropy can be given.
In particular, for the case $\mu = 0$,
the bound (\ref{GenBoussoBound}) is shown to be satisfied 
{\it iff} the cited inequality holds,
and this would mean
that Bousso bound would rest
on the existence of an ultimate lower limit
to the smallness of the scale at which
a thermodynamical description is viable.
This claim, we see, in principle goes beyond the finding
of a condition, physically plausible, 
provably sufficient for the validity of the bound;
it suggests rather that 
the bound itself should be the expression, 
in a general-relativistic context,
of some fundamental property of statistical entropy,
required, as such, simply 
by quantum mechanics and the statistical nature of entropy. 

The scope of the present work is precisely 
to investigate if this is the case,
namely if 
the Bousso
bound and, we could say, the holographic principle,
have some statistical-mechanical flat-spacetime counterpart,
on which they could in principle be ultimately rooted.

We start by revisiting the argument of \cite{Pesci}.
It was shown in it that,
considering a local Rindler horizon
through a point $p$
of a classical spacetime satisfying Einstein equation,
if $\delta A$ is the variation of the cross-sectional area $A$ 
of a pencil of generators of this horizon when an energy flux
comes across the horizon,
for a homogeneous perfect fluid 
under local thermodynamic equilibrium conditions 
Raychaudhuri equation implies

\begin{eqnarray}\label{relate_1}
- \frac{\delta A}{4} = 
\pi l \cdot (\rho + p) l A.
\end{eqnarray}
Here
$l$ is the (small) proper length
covered by the generators in the fluid rest frame
and
$\rho$ and $p$ are 
the local
energy density
(with possible rest-mass energy included) and 
pressure,
respectively.

On the other hand the entropy $dS$
on the null hypersurface $dL$ spanned by the horizon generators
in the same small affine interval
is obviously given by $dS = s l A$,
being $s$ the local entropy density,
so that the generalized Bousso bound (\ref{GenBoussoBound}),
as applied to $dL$, is satisfied if and only if

\begin{eqnarray}\label{microBousso}
s \leq \pi l (\rho + p).
\end{eqnarray}
This equation turns out to be equation (1.9)
of \cite{FlanaganMarolfWald},
reformulated for thin layers,
and also recalls equation (1.11) of the same reference
and equation (3.5) of \cite{BoussoFlanaganMarolf}.
At the end it looks like a sort of local reformulation
of the original bound
$S \leq \pi E D$
(here $D$ is the diameter of the smallest sphere circumscribing the system
and $E$ is energy)
due to Bekenstein \cite{Bekenstein}. 
From (\ref{microBousso}) we get 

\begin{eqnarray}\label{l_star}
l \geq l^* \equiv \frac{1}{\pi} \ \frac{s}{\rho + p}
= \frac{1}{\pi T} \ \left( 1 - \frac{\mu n}{\rho + p} \right),
\end{eqnarray}
where last equality follows from Gibbs-Duhem relation
$\rho = T s - p +\mu n$,
being $T$ and $n$ the local temperature and number density
respectively.
From equation (\ref{microBousso}) note that
Bousso bound demands
the null energy condition be satified
($\rho + p \geq 0$, in present case),
unphysical negative values for
entropy density would be otherwise required.
Equation (\ref{l_star}), in particular when $\mu = 0$, 
recalls
the expression
for the universal bound to
the relaxation timescales
of perturbed systems found 
in \cite{Hod}.\footnote{We are grateful to S. Hod 
for noticing this point.}

At variance with \cite{Pesci},
in deriving equations (\ref{microBousso}) and (\ref{l_star})
no restriction is made on $\mu$. 
The study of the generalized Bousso bound
is focused moreover on
single slices of matter.
We assume that extensivity of entropy means that
total entropy on a collection of regions,
can be expressed as sum of entropies on each component region,
in particular on each component slice of matter,
if we have a slice partition.
When this is true 
we say that, for the chosen partition,
entropy is extensive on each slice;
in this case total entropy cannot depend, of course,
on the partition we have performed.
If we assume that the entropy
which enters the generalized Bousso bound
is extensive in this sense, 
the bound is thus
universally valid if and only if it is
valid for each of such slices.

In principle this connects to the question of
weather it is conceptually possible to have a continuum description
of matter entropy so that total entropy
can be obtained summing on arbitrarily small sub-regions
or, instead, an intrinsic lower limit to the size
of the component sub-regions, in our case to the thickness 
of the slices, must be envisaged.
In this perspective,
equations (\ref{microBousso}) or (\ref{l_star})
say that if we believe in the generalized Bousso
bound, then there is an ultimate lower limit
to the thickness of the slices for which
an extensive notion of entropy can be given,
depending this limit on the assigned thermodynamic conditions.
Conversely, if for 
all
assigned conditions, the 
lower limit (\ref{l_star}) to the thickness of each slice
is supposed to exist, 
then we are sure the generalized Bousso bound will be universally satisfied.

Let us try to look at this in more detail.
Consider a homogeneous macroscopic system $M$
at global thermodynamic equilibrium
and with negligible long-range interactions (such as gravity)
so that its local thermodynamic description is uniform.
Around a point of this system let us imagine to cut
a slice of thickness $l$ large enough that we are sure
it can be considered macroscopic and
the 
statistical-mechanical 
description can be applied
exactly in the same way as for the whole system.
For this slice the notion of entropy is extensive,
in the sense, as above, that joining 
for example two such slices, entropy is doubled.
Now imagine repeating this procedure of the cutting of slices,
with $l$ going lower and lower, but always requiring
the entropy assigned to the slice be extensive.
Well, equation (\ref{l_star})
says that the generalized Bousso bound 
(or the holographic principle) obtains if and only if
there is an ultimate lower limit $l^*$ to $l$,
determined by the thermodynamical state of $M$.  
This argument can evidently be restated in the same way
also for systems only at local thermodynamic equilibrium,
the thermodynamic observables which set the value of $l^*$
being in this case changing from point to point.

This obviously does not mean that, for every system,
$l^*$ is the limiting value $l_{min}$
for which an extensive notion of entropy can be given;
if this would be the case, in fact, 
a slice of every kind of material medium
would saturate Bousso bound,
provided the thickness is $l^*$.
For most systems we can expect $l_{min} \gg l^*$
so that generalized Bousso bound will be satisfied  
by far.

As a 
resum\'e
of what we have said up to now, 
we can state that the generalized Bousso bound
appears to be satisfied if and only if  
the following property obtains:\\
``every
time a system at local thermodynamic equilibrium
has in a point a non-vanishing entropy density,
there exists a lower limit $l^*$ to the spatial scale
on which an extensive notion of entropy can be furnished.
This limit, given by (\ref{l_star}), must be understood as an ultimate
value, impossible to go beyond; for each given system
the actual lower limit $l_{min}$ to the scale for which an extensive notion of
entropy can be given will always satisfy 

\begin{eqnarray}\label{l_min}
l_{min} \geq l^*,
\end{eqnarray}
with in general $l_{min} \gg l^*$.''

This has nothing to do with the mathematical possibility
to perform integrals of entropy density treating this latter
mathematically as a continuous function; this will be always possible
in spite of equation (\ref{l_star}) as extensivity
down to $l=0$ can always be mathematically assumed.
The point is however that
when we require that the {\it mathematical} entropy in a slice
be coincident with the {\it actual} entropy on that slice
as if this slice were alone
(the kind of entropy we expect enters Raychaudhuri equation
and then equation (\ref{relate_1}),
as the focusing of light rays is due, through Einstein equation,
to the {\it actual} energy-momentum they encounter) 
things are different and equation (\ref{l_star}) appears.
In general $l_{min}$ could be expected
to be not lower than local thermalization scale.
With reference to what we said just now however,
the real discriminating point is that the slice be thick enough
that the {\it actual} statistical-mechanical properties we could assign
to it when it is taken alone coincide
with their {\it mathematical} value inferred assuming extensivity.
This implies that $l_{min}$ values also lower than
the local thermalization scale (for example 
lower than the mean free path for an ordinary gas)
could be accepted for consideration, 
if the area $A$ of the slices
can be chosen large enough that a notion of thermodynamic
equilibrium can meaningfully be given also for the slice taken alone
(for the ordinary gas example, a transverse scale
much larger than the mean free path).

We proceed now to study some simple and, in principle, 
well-understood
examples of ideal material media to see what 
$l^*$ and $l_{min}$ are
for them and to verify
if condition (\ref{l_min}) is satisfied.
This will enable us to see which are the mechanisms at work 
that fix $l_{min}$ values satisfying (\ref{l_min})
and then prevent the generalized Bousso bound 
from being
violated.
Fluids at (ultra-)relativistic regimes 
were already considered in \cite{Pesci}.
It has been verified there, in particular, that a photon gas
should satisfy equation (\ref{l_min}), 
with $l^*$ in its form with $\mu =0$. 
A more general argument was also given showing that,
for $\mu \geq 0$, for
material media with constituent particles
with ultra-relativistic dispersion relation,
equation (\ref{l_min}) should be satisfied.  
The argument relied on the statement that
if a slice has thickness smaller than the 
quantum spatial uncertainty
of the constituent particles,
the entropy we could be able to assign to the slice
cannot be extensive. 
This argument seems to imply that the fundamental
mechanism at work to protect Bousso bound could simply
be quantum mechanics, joined to the macroscopic or
statistical nature of entropy.

To investigate this item further,
in the present work we consider 
two non-relativistic systems.
The first is a classical ideal non-relativistic Boltzmann gas.
In this case one could expect that any argument
relying on quantum mechanics should fail,
so that for point-like constituent particles
no lower limit to $l$ is expected to exist,
at least if the saving of extensivity of entropy is concerned.
As a consequence one could infer that
either this classic system can violate Bousso bound
or another mechanism, distinct from quantum mechanics
and thus perhaps of statistical nature alone, 
is also protecting the bound \cite{Gersl}.

The equation of state of a 
non-relativistic ideal Boltzmann gas
with constituent particles of mass $m$,
can be expressed in terms of temperature,
pressure and chemical potential
as 

\begin{eqnarray}\label{state_eq}
\beta p = \frac{z}{\lambda_T^3},
\end{eqnarray}
where, when conventional units are chosen, 
$\beta = 1/ k T$,
$z \equiv e^{\beta (\mu - m c^2)}$
(it is the fugacity,
which turns out to be $z = \lambda_T^3 n$ for a Boltzmann gas)
and $\lambda_T \equiv \sqrt{h^2/ 2 \pi m k T}$.
In the definition of $\lambda_T$, $h$ --in principle
a classically undetermined constant
with the dimensions of an action,
interpreted as the volume of phase space
to be assigned to each degree of freedom--
is the Planck constant 
as, with this choice,
the calculated statistical entropy 
describes correctly
the experimentally verified 
entropy of classical gases
(Sackur-Tetrode equation 
for ideal gases at high temperatures) \cite{Huang}.
Even in a purely classical context thus
the notions of entropy and chemical potential
are in a sense quantum mechanical in that
they require
the introduction of $h$.
$\lambda_T$,
the so-called thermal de Broglie wavelength,
turns out to be roughly
the average de Broglie wavelength for the component particles
and this implies that the complete statistical-mechanical
description of a classical gas requires the introduction
of a sort  of ``effective size'' 
for the component point-like particles,
coincident with their quantum wavelength.
The classical nature of the gas
is entailed in that $z = \lambda_T^3 n \ll 1$;
this assures that the average inter-particle separation
be much larger than $\lambda_T$.
It would desagree with the assumed classical nature
of the gas to choose $l_{min}$ significantly lower than $\lambda_T$.  
On the other hand from the very low value of $z$,
the statistical nature of entropy 
could seem to require
$l_{min} \gg \lambda_T $.
Given however the fact that 
in some circumstances
the area $A$ of the slices
can be chosen in principle even large,
in some cases statistics could in principle be recovered
even when $l_{min} \equiv \lambda_T $ or lower;
this will depend on the actual characteristics
of the systems under consideration.
We see that
this problem with statistics is 
in the end
determined
by the request to consider as point-like,
particles that actually do have an effective quantum size. 
Having at disposition systems
with infinite extension $A$ could solve the statistics problem,
but the quantum mechanical request that $l_{min}$
be not (significantly) lower than $\lambda_T$
cannot be avoided.

From last equality in (\ref{l_star}) and the definition given
for $z$, 
coming back to Planck units we have

\begin{eqnarray}\label{l_star_b}
l^* = \frac{1}{\pi T} \left( 1 - 
\frac{(m + T ln(\lambda_T^3 n)) n}
{(m + g \frac{T}{2}) n + T n} \right) \simeq
\frac{1}{\pi m} \left(
|ln(\lambda_T^3 n)| + \frac{g + 2}{2} \right),
\end{eqnarray}   
where $g$ is the number of degrees of freedom per particle
and the approximation is justified by the
non-relativistic nature of the gas ($\sqrt{T} \ll \sqrt{m}$).
If we parametrize the smallness of $z$
as $\lambda_T^3 n = e^{-\chi}$ with $\chi > 0$,
from (\ref{l_star_b}) and the definition of $\lambda_T$
we get

\begin{eqnarray}\label{comparison_b}
\frac{l^*}{\lambda_T} =
\frac{1}{\sqrt{2 \pi^3}} \sqrt{\frac{T}{m}} \left(
\chi + \frac{g+2}{2} \right).
\end{eqnarray}
Here we see that for $\chi$ not too large,
that is for not unreasonably high rarefaction conditions,
$\chi \ll \sqrt{m/T}$ and thus $l^*/\lambda_T \ll 1$
so that
inequality (\ref{l_min}) is satisfied.    

The lesson we extract from this example is that 
the ideal classical Boltzmann
gas does obey
inequality (\ref{l_min}) and 
the generalized Bousso bound, 
the argument for this appearing 
in most cases statistical 
(depending on the size of
the actual system under consideration),
but, curiously, always quantum mechanical.

The second example we have chosen to consider
is a non-relativistic degenerate ideal Fermi gas.
We assume then density so high that
quantum effects are all important,
namely $\lambda_T^3 n \gg 1$,
and temperature much smaller 
than Fermi energy $\epsilon_{F}$ (still
non-relativistic).
Under these circumstances, number density is
such that the statistical requirement alone
could be always fulfilled also for $l_{min} = \lambda_T$
or smaller; 
extensivity of entropy on the other hand
gets into trouble for slices with thicknesses
significantly smaller than de Broglie wavelength.
$\lambda_T$, 
as evident
in particular in the limit $T \rightarrow 0$,
is however no longer a reliable estimator of the average
de Broglie wavelength of constituent particles.
In any case, calling $\lambda_F$
the de Broglie wavelength at Fermi energy
($\lambda_F = {2 \pi}/\sqrt{2 m \epsilon_F}
= {\pi \sqrt{2}}/\sqrt{m \epsilon_F}$),
under degeneracy conditions
$l_{min}$
cannot be significantly lower than $\lambda_F$. 

Now, as can be inferred from Gibbs-Duhem relation
for degenerate Fermi gases
(see for example \cite{Huang}), 

\begin{eqnarray}
s = n \frac{\pi^2}{2} \left( \frac{T}{\epsilon_F} \right)
\end{eqnarray}
at first order in $T/\epsilon_F$,
so that from equation (\ref{l_star}),
neglecting higher order terms in $T/\epsilon_F$,
we get

\begin{eqnarray}\label{l_star_fermi}
l^* = \frac{\pi}{2} \ \frac{T}{\epsilon_F m + \epsilon_F^2}.
\end{eqnarray}
Finally, 
using also $\sqrt{\epsilon_F} \ll \sqrt{m}$ (non-relativistic conditions),  
we have

\begin{eqnarray}
l^* \ll
\frac{\pi}{2} \ \frac{1}{m} \ll
\pi \sqrt{2} \ \frac{1}{\sqrt{m \epsilon_F}} =
\lambda_F
\end{eqnarray}
and this implies equation (\ref{l_min}) is satisfied.
In this second example it is thus always the request
of the saving of extensivity of entropy 
that
prevents
Bousso bound to be violated.
From (\ref{l_star_fermi}), when $T = 0$ we get
$l^* = 0$ ($s = 0$) so that
the generalized Bousso
bound is satisfied trivially.

At the end 
the question of how $l_{min}$ could precisely
be defined as a property of the given material medium
(i.e. irrespective of the size of the actual system under consideration;
assuming that systems large enough can be considered,
to overcome possible problems with statistics) 
seems to be answered
through the following criterium:
$l_{min}$ be determined by the requirement 
that 
$l_{min} \ge \Delta l$, 
with $\Delta l$
the quantum spatial uncertainty of the constituents
(or the mimimum between this
and their size if they are composite objects).
In this regard
note that 
the order-of-magnitude estimate of $l_{min}$
for a photon gas given in \cite{Pesci},
suggesting that the inequality 
$l_{min} \ge l^*$ was broadly respected,
gave anyway a value for $l_{min}$ remarkably close to $l^*$.
When, 
with reference to the definition above for $l_{min}$,
we try to perform a bit more careful estimation
we find that $l_{min}$ for a photon gas is like to be
practically coincident with $l^*$. 
If, in fact, 
we assume that $\Delta\epsilon = \epsilon/2$
(being $\epsilon = 2.82 T$ the peak energy \cite{Kittel})
does roughly capture the maximal uncertainty
in photon energy compatible 
with the given Planck's law at temperature $T$,
from time-energy uncertainty relation 
$\Delta t \Delta\epsilon \ge \frac{1}{2}$ we obtain
$\Delta t \ge \frac{1}{\epsilon}$ so that
$\Delta l = \frac{1}{\epsilon} = \frac{1}{2.82 T}$.
This value is very close to the limiting scale $l^*$,
which in this case is $l^* = \frac{1}{\pi T}$.    
The entropy of this slice
--constructed in such a way that the restriction
on the spatial coordinate imposed by slicing
has a value
with limiting compatibility with the request to not destroy
the thermodynamical features of the gas--
practically saturates the generalized Bousso bound.
We find this a quite strong indication that 
thermodynamically-meaningful slices 
of photon gases
(for sure, among the most entropic media)
can be defined down to scales small enough
to be able to saturate the bound,
with extensivity of entropy progressively lost
when going below.

Let us conclude with some 
remarks on the results we have obtained.
The covariant entropy bound appears
to be expression of, or equivalent to,
the existence of a fundamental lower limit
to the scale of the thermodynamic description.
We see that from something living in principle
in general relativity we have come 
to something else
linked to concepts of statistical mechanics
or thermodynamics on flat spacetime.
Inequality (\ref{microBousso}) corresponds
to a limit on entropy for $\rho$ and $p$ assigned.
In flat spacetime however this does not mean any intrinsic limit
on the total entropy inside an assigned volume,
as $s$ can grow without limit with $\rho$ and $p$.
When gravity is turned on,
the same limit (\ref{microBousso}) on entropy density 
acquires an additional meaning
in terms of the gravitational focusing
accompanying the given $\rho$ and $p$
and this brings to that now
for an assigned geometry, let say on an assigned lightsheet,
an absolute intrinsic limit to entropy can be envisaged.
A fundamental statistical-mechanical property,
the existence of a limiting length,
acquires through gravity a fundamental additional valence
as absolute limit on entropy.

The role of general relativity is then to shape
the peculiar form the generalized Bousso 
bound has and, in particular, 
under suitable circumstances
the prodigious relation (\ref{spherical}).
The entropy limit, being determined by gravitational lensing,
increases when gravity decreases;
without gravity no entropy limit at all could be expected.
The setting of this limit however would not be possible
if an intrinsic lower limit $l^*$ would not exist
to the spatial scale for which
a statistical-mechanical description is viable. 
Gravity succeeds
in determining the absolute entropy limit, 
only thanks to the statistical-mechanical relations
(\ref{microBousso}) and (\ref{l_star}).

Equations (\ref{microBousso}) and (\ref{l_star})
do not depend on gravity.
In fact, even if our derivation of them
hinges on the gravitational lensing
of light rays to produce the right cross-sectional area variation 
required by Einstein equation,
also the expression of an assigned entropy as an area
depends in the same manner on the strength of gravitational interaction
so that this latter does not enter the game.
Relations (\ref{microBousso}) and (\ref{l_star}) 
live in statistical mechanics
and should be in principle universally provable 
through strictly statistical-mechanical
arguments; we have shown above that
a photon gas seemingly permits to choose
slices of limiting thickness small enough
to saturate them.
The possible (false) impression they could depend on gravity
is perhaps generated simply by our path
to them, 
namely we arrived at them from a general relativistic scenario:
the generalized Bousso bound.
Experimentally-viable gravity theories
different from general relativity,
should bring to the same estimation (\ref{l_star}) 
for the limiting length.

Note that $l^*$ in (\ref{l_star}) can be rewritten
in conventional units as

\begin{eqnarray}\label{conv}
l^*_{conv} = \frac{1}{\pi} \ \frac{c \hbar}{k} \ \frac{s}{\rho + p} = 
\frac{1}{\pi} \ l_{Planck} \ \frac{T_{Planck}}{T} \ \frac{T s}{\rho + p}
\end{eqnarray}
being $c$ the speed of light in vacuum, $\hbar$  
the (reduced) Planck constant and $k$ the Boltzmann constant and
$l_{Planck}$ and $T_{Planck}$ the Planck length and temperature.
In the first equality here, the quantum origin of $l^*$ is manifest.
The second equality shows that in general 
$l^*_{conv} \gg l_{Planck}$, 
so that matter entropy
can saturate Bousso bound
on scales much larger than Planck scale,
even if usually this is not the case
(a recent example is in \cite{Gersl2}),
just because of 
the geometry of the lightsheet considered
or of the properties
of material medium (in general $l_{min} \gg l^*$).
On the other hand, as said, 
thermodynamical conditions can also be such that $l^*_{conv} = 0$
and in this case Bousso bound can never be challenged
(the Fermi gas example above).
In general putting
$\frac{T s}{\rho + p} \equiv \eta$,
for systems for which $\eta = {\cal O}(1)$
we obtain $l^*_{conv} =  {\cal O}(l_{Planck})$
when $T = {\cal O}(T_{Planck})$.

In equations (\ref{conv}), $l^*$ does not depend on $G$,
the Newton constant; in particular it retains its value also
when $G = 0$. 
This is expected if this limiting length $l^*$  
from which the generalized
Bousso bound arises has no relation with gravity and
it is then intrinsically statistical mechanical.
What does depend on $G$ is the maximum entropy
on a lightsheet, namely the actual limiting entropy value
entering the generalized Bousso bound,
going this maximum as $1/G$.

The generalized Bousso bound implies
(provided the ordinary second law is assumed to hold)
the generalized second law of thermodynamics \cite{FlanaganMarolfWald}.
When the areas $A$ and $A^\prime$ entering
the generalized Bousso bound are black hole areas,
$A/4$ and $A^\prime/4$ are black hole entropies.
From our results
we see that
statistical mechanics and general relativity
imply the generalized Bousso bound to hold, 
with some conventional fluids
(the photon gas, for example) 
seemingly able to saturate it,
for thin layers but 
in any case at a scale very far from the Planck length.
We deduce from this that, 
if to black holes an entropy supposedly
entering a second law can be assigned,
the fact that its value is $A/4$ appears to be required 
merely by (conventional) statistical mechanics and general relativity.
In other words,
always a same thing, black hole entropy,
is what is produced using the same ingredients,
quantum mechanics plus general relativity;
in our case quantum mechanics acts
cospiring in order that
statistical mechanics
respects the lower limiting scale $l^*$.
It is like to have two equivalent ways 
of looking at black hole entropy:
Bekenstein-Hawking approach
\cite{Bekenstein, Hawking}
(black hole entropy from the point of view of the vacuum) 
and the present statistical-mechanical approach
to Bousso bound
(black hole entropy from the point of view of material media).

The saturation of Bousso bound,
corresponding roughly to about 1 bit of information
per Planck area, points, as such, to
Planck-scale physics. 
This limit however, 
demands for
a fundamental
property of 
flat-spacetime
statistical mechanics, 
namely the existence
of the lower limiting scale $l^*$, 
not affected by
Planck-scale physics.
This fact seems
to agree with the expectation that
this limiting value of entropy and,
in particular, 
the bulk of black hole entropy
(or of Hawking radiation)
are not related to Planck-scale physics
(recent results in this direction for the case 
of acceleration radiation are in \cite{Rinaldi_Navarro}).

I thank Jan Ger\v{s}l for the remarks on an earlier
version of this manuscript, in particular for
the correction of an error and for
an important point he has raised (besides \cite{Gersl}).

\end{document}